\documentclass[prd,12pt,
tightenlines,preprintnumbers,showpacs,%
superscriptaddress,nofootinbib,amsmath,amssymb]{revtex4}

\begin{document}

\preprint{CECS-PHY-05-01} \preprint{hep-th/0501040}

\title{Scalar Fields Nonminimally Coupled to \emph{pp} Waves}

\author{Eloy Ay\'on-Beato }\email{ayon-at-cecs.cl}
\affiliation{Centro~de~Estudios~Cient\'{\i}ficos~(CECS),%
~Casilla~1469,~Valdivia,~Chile.}
\affiliation{Departamento~de~F\'{\i}sica,~CINVESTAV--IPN,%
~Apdo.~Postal~14--740,~07000,~M\'exico~D.F.,~M\'exico.}
\author{Mokhtar Hassa\"{\i}ne}\email{hassaine-at-inst.mat.utalca.cl}
\affiliation{Instituto de
Matem\'atica y F\'{\i}sica, Universidad de Talca, Casilla 747,
Talca, Chile.}
\affiliation{Centro~de~Estudios~Cient\'{\i}ficos~(CECS),%
~Casilla~1469,~Valdivia,~Chile.}

\begin{abstract}
Here, we report \emph{pp} waves configurations of three-dimensional
gravity for which a scalar field nonminimally coupled to them acts
as a source. In absence of self-interaction the solutions are
gravitational plane waves with a profile fixed in terms of the
scalar wave. In the self-interacting case, only power-law potentials
parameterized by the nonminimal coupling constant are allowed by the
field equations. In contrast with the free case the self-interacting
scalar field does not behave like a wave since it depends only on
the wave-front coordinate. We address the same problem when
gravitation is governed by topologically massive gravity and the
source is a free scalar field. From the \emph{pp} waves derived in
this case, we obtain at the zero topological mass limit, new
\emph{pp} wave solutions of conformal gravity for any arbitrary
value of the nonminimal coupling parameter. Finally, we extend these
solutions to the self-interacting case of conformal gravity.
\end{abstract}


\maketitle

\section{Introduction}

Recently, it has been shown that a version of three-dimensional
gravity governed just by the Cotton tensor with a conformally
invariant free source given by a scalar field admits \emph{pp} wave
solutions \cite{Deser:2004wd}. The same model with a conformal
self-interacting potential also supports such gravitational fields
\cite{Ayon-Beato:2004fq}. In view of these two works, a natural
question is to ask whether \emph{pp} wave configurations are also
solutions to standard three-dimensional Einstein gravity with scalar
source. This problem is of interest since it is well known that pure
gravitational waves are forbidden in three dimensions. In contrast
with these previous works, since the Einstein tensor is not
conformally invariant, this allows to consider a scalar field
nonminimally coupled to gravity for which the coupling parameter is
not necessarily the conformal one. The three-dimensional action we
consider here is given by
\begin{equation}
S(g_{\alpha\beta},\Phi)=\int d^3x\,\sqrt{-g}\left(\frac{1}{2\kappa}R
-\frac{1}{2}\nabla_{\alpha}\Phi\nabla^{\alpha}\Phi-\frac{1}{2}\xi
R\,\Phi^2-U(\Phi)\right), \label{eq:action}
\end{equation}
where $\xi$ is the nonminimal coupling parameter and $U(\Phi)$ is
the self-interaction potential. The field equations obtained by
varying the metric (resp.\ the scalar field) read
\begin{equation}
G_{\alpha\beta}={\kappa}T_{\alpha\beta}, \label{eq:Einsteineqs}
\end{equation}
and
\begin{equation}
\Box\Phi=\xi R\,\Phi+\frac{\mathrm{d}U(\Phi)}{\mathrm{d}\Phi},
\label{eq:waveeq}
\end{equation}
where the energy-momentum tensor is given by
\begin{equation}
T_{\alpha\beta}=\nabla_{\alpha}\Phi\nabla_{\beta}\Phi
-g_{\alpha\beta}\left(\frac{1}{2}\nabla_{\sigma}\Phi\nabla^{\sigma}\Phi
+U(\Phi)\right)
+\xi\left(g_{\alpha\beta}\Box-\nabla_{\alpha}\nabla_{\beta}
+G_{\alpha\beta}\right)\Phi^2. \label{eq:nrg}
\end{equation}

In what follows we shall explore the existence of \emph{pp} wave
configurations for the previous system. In the next section, we
present the independent field equations for a \emph{pp} wave
background in three dimensions. Sec.~\ref{sec:U=0} is devoted to the
free case (i.e., $U(\Phi)=0$) for which it is shown that the
solutions are gravitational plane waves whose profiles are
determined by the wave profile of the free scalar field. The
analysis of the self-interacting case is done in
Sec.~\ref{sec:U<>0}.  In this set-up, only power-law potentials are
allowed by the field equations where the power is given in terms of
the nonminimal coupling parameter. In contrast with the free
situation, the scalar field does not behave like a wave in
accordance with its self-interacting nature. Moreover, for a
nonminimal coupling parameter $\xi=1/2$, the potential reduces to a
positive constant which allows to interpret the solutions as
\emph{pp} waves for a free scalar field when the Einstein equations
are supplemented by a positive cosmological constant. In
Sec.~\ref{sec:TMG}, we present the \emph{pp} wave solutions of the
topologically massive gravity when the scalar field is free. The
related gravitational waves exhibit an effective mass given in terms
of the nonminimal coupling parameter. In Sec.~\ref{sec:CG}, we show
that at the zero topological mass limit, the above solutions turn
out to be \emph{pp} wave solutions of conformal gravity for any
arbitrary value of the nonminimal coupling parameter. This result is
a generalization of the conformal \emph{pp} waves of
Ref.~\cite{Deser:2004wd} since the energy-momentum tensor of these
solutions evaluated on-shell is traceless independently of the value
of the nonminimal coupling parameter. Taking into consideration this
last remark, we have also derived in the same section, the \emph{pp}
wave configurations of conformal gravity with a self-interacting
nonminimally scalar field. Finally, the appendices are devoted to
the analysis of some special values of the nonminimal coupling
parameter as for example the case $\xi=1/4$ which requires a
separate derivation and for which the allowed configuration
corresponds to a free massive scalar field.

\section{\label{sec:pp}\emph{pp} wave field equations}

The term \emph{pp} wave is an abbreviation for plane-fronted
gravitational waves with parallel rays, which are the gravitational
configurations possessing a covariantly constant null vector field
\cite{Ehlers:1962}. The corresponding geometry is written in three
dimensions as
\begin{equation}
ds^2=-F(u,y)du^2-2dudv+dy^2, \label{eq:ppwave}
\end{equation}
where the covariantly constant null vector field is $\partial_v$. In
three dimensions, the front of the wave (the surfaces at constant
$u$, $v$) is just a line and not a plane as it occurs in higher
dimensions, and hence it would be more appropriate to call the
geometries (\ref{eq:ppwave}) line fronted gravitational waves.
However, in order to avoid any confusions, we shall use the term
\emph{pp} wave as usual.

The null field $\partial_v$ is a Killing field and so we impose the
same symmetry on the source, that means $\Phi=\Phi(u,y)$. For the
geometry (\ref{eq:ppwave}), the only nonvanishing component of the
Einstein tensor is $G_{uu}$. Consequently, all the components of the
energy-momentum tensor except $T_{uu}$ must vanish by virtue of the
Einstein equations. For convenience, we choose the following
combinations of this tensor
\begin{eqnarray}
\label{eq:uv+yy} T_{uv}+T_{yy}&=&(1-2\xi)(\partial_y\Phi)^2
                -2\xi\Phi\partial^2_{yy}\Phi=0, \\
\label{eq:uy} T_{uy}       &=&(1-2\xi)\partial_u\Phi\partial_y\Phi
                -2\xi\Phi\partial^2_{uy}\Phi=0, \\
\label{eq:yy} T_{yy}       &=&\frac12(\partial_y\Phi)^2-U(\Phi)=0,
\end{eqnarray}
while the remaining independent Einstein equation can be taken as
\begin{equation}\label{eq:uu-uv}
G_{uu}-\kappa(T_{uu}-FT_{uv})=\frac12(1-\kappa\xi\Phi^2)\partial^2_{yy}F
-\kappa\xi\Phi(\partial_y\Phi\partial_yF-2\partial^2_{uu}\Phi)
-\kappa(1-2\xi)(\partial_u\Phi)^2=0.
\end{equation}
The system of Eqs.~(\ref{eq:uv+yy})-(\ref{eq:uu-uv}) are all the
independent Einstein equations and in what follows we shall always
refer to these equations. We do not take into consideration the
scalar equation (\ref{eq:waveeq}) which takes the form
\begin{equation}\label{eq:waveeqpp}
\partial^2_{yy}\Phi=\frac{\mathrm{d}U(\Phi)}{\mathrm{d}\Phi},
\end{equation}
since the conservation of the energy-momentum tensor (\ref{eq:nrg})
together with the existence of a nontrivial scalar field solution of
the Einstein equations (\ref{eq:Einsteineqs}) guarantee that this
equation is satisfied.

In the next section, we analyze the free case, $U(\Phi)=0$, for
which the solutions turn out to be plane waves, i.e., where the
metric dependence on the front-wave coordinate $y$ is just
quadratic.

\section{\label{sec:U=0}Free scalar fields: plane waves}

In this section, we consider a scalar field nonminimally coupled to
gravity without self-interaction (i.e., $U(\Phi)=0$). In particular,
we include as a first study the minimal coupling $\xi=0$ since the
fields equations do not allow to consider a potential in this case.

For a scalar field minimally coupled to gravity (i.e., $\xi=0$), the
Einstein equations become of first order for the scalar field. In
particular, combining Eqs.~(\ref{eq:uv+yy}) and (\ref{eq:yy}) we
obtain
\begin{eqnarray}
T_{uv}+T_{yy}&=&(\partial_y\Phi)^2=0, \\
T_{uv}-T_{yy}&=&2U(\Phi)=0,
\end{eqnarray}
and conclude that the self-interaction must be absent while the
scalar field depends only on the retarded time, $\Phi=\Phi(u)$. The
remaining equation (\ref{eq:uu-uv}) now reduces to
\begin{equation}
\frac12\partial^2_{yy}F=\kappa\left(\frac{\mathrm{d}\Phi}
{\mathrm{d}u}\right)^2,
\end{equation}
and integrates as
\begin{equation}\label{eq:solF0}
F(u,y)=\kappa\left(\frac{\mathrm{d}\Phi}{\mathrm{d}u}\right)^2y^2
+F_1(u)y+F_0(u),
\end{equation}
where $F_1$ and $F_0$ are two integration functions. It is
well-known that any dependence up to first grade on the front-wave
coordinate $y$ in $F$ can be eliminated through a coordinate
transformation \cite{Ehlers:1962}. For example, for a generic
dependence
\begin{equation}\label{eq:gpw}
F(u,y)=F_2(u)y^2+F_1(u)y+F_0(u),
\end{equation}
the corresponding transformation which permits to eliminate $F_1$
and $F_0$ is
\begin{equation}\label{eq:coordpw}
\textstyle(u,v,y)\mapsto\left(u,
v-\frac12\frac{\mathrm{d}B}{\mathrm{d}u}(2y-B)
+\frac14\int\mathrm{d}u(F_1B+2F_0),y-B\right),
\end{equation}
where $B=B(u)$ is a function satisfying the linear equation
\[
\frac{\mathrm{d}^2B}{\mathrm{d}u^2}+F_2B=-\frac12F_1.
\]
Applying the coordinate transformation (\ref{eq:coordpw}) to our
case we obtain the following solution
\begin{subequations}\label{eq:minsol}
\begin{eqnarray}
ds^2&=&-\kappa\left(\frac{\mathrm{d}\Phi}{\mathrm{d}u}\right)^2
y^2du^2-2dudv+dy^2,\\
\Phi&=&\Phi(u),
\end{eqnarray}
\end{subequations}
corresponding to the geometry of a plane wave with profile fixed by
the scalar field which depends arbitrarily on the retarded time.

A natural question following from this analysis is to ask whether
such free scalar field configurations still exist if one includes a
nonminimal coupling to gravity. The procedure is similar to what we
done for the minimal case. Indeed, the absence of potential together
with Eq.~(\ref{eq:yy}) imply
\[
\partial_y\Phi=0,
\]
from which we conclude again that the scalar field depends only on
the retarded time and, hence, Eqs.~(\ref{eq:uv+yy}) and
(\ref{eq:uy}) are trivially satisfied. The remaining independent
Einstein equation (\ref{eq:uu-uv}) becomes now
\begin{equation}
\frac12\partial^2_{yy}F=\frac{\kappa}{1-\kappa\xi\Phi^2}
\left[(1-2\xi)\left(\frac{\mathrm{d}\Phi}{\mathrm{d}u}\right)^2-2\xi\Phi
\frac{\mathrm{d}^2\Phi}{\mathrm{d}u^2}\right],
\end{equation}
and the absence of dependence on $y$ in the right hand side gives
straightforwardly
\begin{equation}
F(u,y)= \frac{\kappa}{1-\kappa\xi\Phi^2}
\left[(1-2\xi)\left(\frac{\mathrm{d}\Phi}{\mathrm{d}u}\right)^2-2\xi\Phi
\frac{\mathrm{d}^2\Phi}{\mathrm{d}u^2}\right]y^2+F_1(u)y+F_0(u).
\end{equation}
As done previously, the coordinate transformation (\ref{eq:coordpw})
allows to eliminate the functions $F_1$ and $F_0$. Thus, the
solution in the free case corresponds to a plane wave with its
profile determined again from the wave profile of the scalar field
\begin{subequations}\label{eq:freesol}
\begin{eqnarray}
ds^2&=&-\frac{\kappa}{1-\kappa\xi\Phi^2}
\left[(1-2\xi)\left(\frac{\mathrm{d}\Phi}{\mathrm{d}u}\right)^2-2\xi\Phi
\frac{\mathrm{d}^2\Phi}{\mathrm{d}u^2}\right]y^2du^2-2dudv+dy^2,\\
\Phi&=&\Phi(u).
\end{eqnarray}
\end{subequations}
It is interesting to note that the value $\xi=0$ which corresponds
to the minimal coupling is not singular in the expression
(\ref{eq:freesol}) and yields precisely to the minimal solution
(\ref{eq:minsol}).

We conclude that in absence of potential, the system of equations
(\ref{eq:Einsteineqs}-\ref{eq:waveeq}) supports plane wave
gravitational fields for which the source also behaves like a wave.
In the next section, we show that the introduction of a potential
whose form is dictated by the field equations breaks the wavy
behavior of the scalar field while the geometries are now
interpreted as \emph{pp} waves.

\section{\label{sec:U<>0}Self-interacting scalar fields: \emph{pp} waves}

We now consider a scalar field nonminimally coupled to gravity with
a self-interaction potential. In order to achieve this analysis, it
is judicious to make the following redefinition
\begin{equation}\label{eq:Phi2sigma}
\Phi=\frac1{\sigma^{2\xi/(1-4\xi)}},
\end{equation}
which obviously excludes the value $\xi=1/4$. This case deserves a
separate analysis since Eqs.~(\ref{eq:uv+yy}) and (\ref{eq:uy})
integrate as logarithms for this specific value (see Appendix
\ref{app:xi=1/4}). The equations (\ref{eq:uv+yy}-\ref{eq:uy})
expressed in terms of $\sigma$ become
\begin{subequations}\label{eq:s_yyuy}
\begin{eqnarray}
\partial^2_{yy}\sigma &=& 0, \\
\partial^2_{uy}\sigma &=& 0,
\end{eqnarray}
\end{subequations}
which implies that $\sigma$ is separable in $u$ and $y$ and linear
in $y$, i.e.,
\begin{equation}\label{eq:subssigma}
\sigma(u,y)=2\sqrt{\lambda}y+f(u),
\end{equation}
where $\lambda$ is a positive constant and $f$ is an undetermined
function of the retarded time. Inserting the above expression into
Eq.~(\ref{eq:yy}) imposes the self-interaction potential to be of
the form
\begin{equation}\label{eq:U(Phi)}
U_\xi(\Phi)=\frac{8\xi^2\lambda}{(1-4\xi)^2}\Phi^{(1-2\xi)/\xi}.
\end{equation}
We would like to stress that the emergence of such potential is
interesting. Indeed, as it is well-known, a scalar field conformally
coupled to $D-$dimensional gravity requires the nonminimal coupling
parameter to be chosen as $\xi=\xi_D=(D-2)/[4(D-1)]$. Surprisingly,
the allowed self-interaction potential which does not spoil the
conformal invariance of the scalar field is precisely the one
obtained here (\ref{eq:U(Phi)}) when it is written in terms of the
corresponding conformal coupling. For example in three dimensions,
the conformal coupling is $\xi=1/8$ and so the potential
(\ref{eq:U(Phi)}) becomes
\begin{equation}\label{eq:U1/8(Phi)}
U_{1/8}(\Phi)=\frac{\lambda}{2}\Phi^6,
\end{equation}
which corresponds to the conformally invariant potential in three
dimensions. Another interesting value is $\xi=1/2$ for which the
potential (\ref{eq:U(Phi)}) becomes a positive constant. Hence,
taking the value $\xi=1/2$ is equivalent to consider the Einstein
equations with an effective positive cosmological constant of value
\begin{equation}\label{eq:Lambda}
\Lambda={\kappa}U_{1/2}(\Phi)=2\kappa\lambda,
\end{equation}
and without self-interaction potential.

We now go back to our analysis for which we have fully determined
the scalar source. As a simple check, it can be shown that the
scalar field given by Eqs.~(\ref{eq:Phi2sigma}) and
(\ref{eq:subssigma}) is a solution of the nonlinear Klein-Gordon
equation on a \emph{pp} wave background (\ref{eq:waveeqpp}) for the
potential (\ref{eq:U(Phi)}). Hence, it remains to obtain the metric
function $F$ from Eq.~(\ref{eq:uu-uv}). In order to achieve this
task, it is convenient to define new independent variable and
function as
\begin{subequations}\label{eq:xH2yF}
\begin{eqnarray}
x&=&\kappa\xi(2\sqrt{\lambda}y+f)^{-4\xi/(1-4\xi)}, \\
H(u,x)&=& \frac{F}{2\sqrt{\lambda}y+f}-\frac{1}{2\lambda}
\frac{\mathrm{d}^2f}{\mathrm{d}u^2}.
\end{eqnarray}
\end{subequations}
With the above redefinitions, Eq.~(\ref{eq:uu-uv}) is now written as
\begin{equation}\label{eq:hypergeomDE}
x(x-1)\partial^2_{xx}H+\frac{(12\xi-1)x+1-8\xi}{4\xi}\partial_xH
-\frac{1-4\xi}{4\xi}H=0,
\end{equation}
from which we recognize the hypergeometric differential equation
\cite{Erdelyi:1953}. The general solution of equation
(\ref{eq:hypergeomDE}) is
\begin{equation}\label{eq:solH}
H(u,x)=F_1(u)\,_2\!\tilde{F}_1\!\left(1,\frac{4\xi-1}{4\xi};
                              \frac{8\xi-1}{4\xi};x\right)
      +F_2(u)\left(\frac{x}{\kappa\xi}\right)^{(1-4\xi)/(4\xi)},
\end{equation}
where $F_1$ and $F_2$ are integration functions, and
$_2\!\tilde{F}_1(a,b;c;x)$ denotes the hypergeometric function with
parameters $a$, $b$, and $c$. Obviously, the above representation in
terms of the hypergeometric function is only valid when the
hypergeometric function is well-defined. As it is shown in details
in Appendix \ref{sec:xi_n}, the nonminimal coupling values
$\xi_n=1/[4(2+n)]$, $n=0,1,2,\ldots$, are excluded for this reason.
Their corresponding \emph{pp} wave configurations will be analyzed
separately in this Appendix. For the other values we evaluate the
solution (\ref{eq:solH}) in the original variables (\ref{eq:xH2yF})
and we perform the following coordinate change, which allows to
eliminate from the metric up to the first grade dependence on the
front-wave coordinate $y$,
\begin{equation}\label{eq:coordppw}
\textstyle(u,v,y)\mapsto\left(u,
v+\frac{1}{4\lambda}\left({\mathrm{d}f}/{\mathrm{d}u}\right)
(2\sqrt{\lambda}y+f) -\frac{1}{8\lambda}\int\mathrm{d}u
\left[\left({\mathrm{d}f}/{\mathrm{d}u}\right)^2
-4\lambda{F_2}\right],y+\frac{f}{2\sqrt{\lambda}}\right).
\end{equation}
This transformation is equivalent to put $f=0$ and $F_2=0$ and
hence, it clearly shows that the undetermined dependence on the
retarded time of the scalar field (\ref{eq:subssigma}) can be
removed. This fact is an obvious consequence of the self-interacting
character of the source. Finally, for a generic nonminimal coupling
parameter the solution is given by
\begin{subequations}\label{eq:sol}
\begin{eqnarray}
ds^2&=&-F_1(u)\,_2\!\tilde{F}_1\!\left(1,\frac{4\xi-1}{4\xi};
                               \frac{8\xi-1}{4\xi};
       \kappa\xi\Phi^2\right)
       2\sqrt{\lambda}ydu^2-2dudv+dy^2,\qquad~\\
\Phi&=&(2\sqrt{\lambda}y)^{-2\xi/(1-4\xi)}.
\end{eqnarray}
\end{subequations}
More precisely, for $\xi\not=1/2$, the previous solution is
interpreted as a \emph{pp} wave for a self-interacting scalar field
nonminimally coupled to gravity. For the value $\xi=1/2$, the
solution (\ref{eq:sol}) is well defined and can be expressed in
terms of the effective cosmological constant (\ref{eq:Lambda}) as
\begin{subequations}\label{eq:sol1/2}
\begin{eqnarray}
ds^2&=&-F_1(u)\mathrm{arctanh}(\sqrt{\Lambda}y)du^2-2dudv+dy^2,\\
\Phi&=&\sqrt{\frac{2\Lambda}{\kappa}}\,y,
\end{eqnarray}
\end{subequations}
and can be seen as a solution for a free scalar field nonminimally
coupled to gravity (with parameter $\xi=1/2$) in presence of a
positive cosmological constant. It is interesting to note that in
contrast with the free solutions of Sec.~\ref{sec:U=0}, where the
cosmological constant is absent, its introduction breaks the wave
behavior of the scalar field, inducing a linear dependence on front
coordinate and, changes the plane wave character of the
gravitational field.

This last remarks opens naturally the discussion about the
introduction of a cosmological constant in our original system of
equations. It is simple to see that in the free case, the only
possibility occurs for the nonminimal coupling parameter value
$\xi=1/2$ and for a positive cosmological constant. The solution is
precisely the one obtained in Eqs.~(\ref{eq:sol1/2}). In the
self-interacting case the introduction of a cosmological constant is
trivial. It is equivalent of having an effective potential
\[
U_{\mathrm{eff}}(\Phi)=U(\Phi)+\frac{\Lambda}{\kappa},
\]
without cosmological constant, and hence the results of this section
applies for this effective potential. Returning to the original
potential this is equivalent to add the cosmological term to both
sides of the Einstein equations.

This analysis completes the study of scalar fields (including or not
a self-interaction) nonminimally coupled to \emph{pp} waves, when
gravitation is described by the standard Einstein tensor. In the
next section we address the same problem when gravitation is
governed by topologically massive gravity for which the Einstein
tensor is supplemented by the Cotton tensor. For simplicity, this
analysis is done only in the free case.

\section{\label{sec:TMG}Topologically Massive Gravity \emph{pp} waves}

In this section we extend the scope of this work to topologically
massive gravity which is an alternative gravitational theory in
three dimensions introduced by Deser, Jackiw, and Templeton
\cite{Deser:1981wh}. This theory is obtained by adding to the $2+1$
gravity action (\ref{eq:action}) the topological Chern-Simons term
for the local Lorentz group. The corresponding field equations
obtained by varying the metric read
\begin{equation}\label{eq:TMG}
\frac1{\mu}C_{\alpha\beta}+G_{\alpha\beta}={\kappa}T_{\alpha\beta},
\end{equation}
where $\mu$ is the topological mass, $C_{\alpha\beta}$ is the
symmetric, conserved, and traceless Cotton tensor defined by
\begin{equation}\label{eq:Cotton}
C^{\alpha\beta}=\frac{1}{\sqrt{-g}}\epsilon^{\alpha\sigma\nu}
D_{\sigma}\left(R_{\nu}^{~\beta}
-\frac{1}{4}\delta_{\nu}^{~\beta}R\right),
\end{equation}
while the energy-momentum is given by Eq.~(\ref{eq:nrg}). Note that
for lightlike sources, the \emph{pp} wave geometries of this theory
have been previously considered in Refs.~\cite{Deser:1992nk}

For later convenience, we first consider the vacuum case of
equations (\ref{eq:TMG}) for a \emph{pp} wave background. As it is
well-known, one of the most important differences between standard
$2+1$ gravity and topologically massive gravity is that, this later
supports gravitational waves in vacuum since it has at least one
propagating degree of freedom. For the \emph{pp} wave geometry, the
only nonvanishing component of the Cotton tensor is
$C_{uu}=\frac{1}{2}\partial^3_{yyy}F$, and consequently in absence
of sources, the only nonvanishing equation is
\begin{equation}\label{eq:TMGveq}
\frac1{\mu}C_{uu}+G_{uu}=\frac{1}{2\mu}\partial^3_{yyy}F
+\frac{1}{2}\partial^2_{yy}F=0.
\end{equation}
This equation is integrated as
\begin{equation}\label{eq:TMGvsol}
F(u,y)=F_2(u)\mathrm{e}^{-\mu{y}}+F_1(u)y+F_0(u),
\end{equation}
and, after eliminating up to the first grade dependence on $y$,
gives the following \emph{pp} wave geometry
\begin{equation}\label{eq:TMGppv}
ds^2=-F_2(u)\mathrm{e}^{-\mu{y}}du^2-2dudv+dy^2.
\end{equation}
It is easy to see that the metric function $F$ satisfies the linear
Klein-Gordon equation
\begin{equation}\label{eq:K_Gmu}
\Box{F}=\mu^2F,
\end{equation}
with mass $\mu$ which justifies the massive character of the theory.

We now consider the same problem with the scalar source. Since the
Cotton tensor involves third-order derivatives of the metric
function $F$, we restrict ourselves to the free case (i.e., without
a self-interacting potential). As seen previously, the Cotton tensor
has only a contribution along the component $uu$ of the equations
(\ref{eq:TMG}) and hence the analysis done in Sec.~\ref{sec:U=0} for
the free case is still valid for the other components of the
equations. This means that the scalar field depends only on the
retarded time $\Phi=\Phi(u)$, while the $uu$ equation becomes now
\begin{equation}\label{eq:TMGuu}
\frac{1}{2\mu}\partial^3_{yyy}F
+\frac12(1-\kappa\xi\Phi^2)\partial^2_{yy}F
=\kappa\left[(1-2\xi)\left(\frac{\mathrm{d}\Phi}{\mathrm{d}u}\right)^2
-2\xi\Phi\frac{\mathrm{d}^2\Phi}{\mathrm{d}u^2}\right].
\end{equation}
The integration of this equation yields to the following solution
\begin{eqnarray}
F(u,y)&=&F_2(u)\exp\!\left[-\mu(1-\kappa\xi\Phi^2)y\right]
+\frac{\kappa}{1-\kappa\xi\Phi^2}
\left[(1-2\xi)\left(\frac{\mathrm{d}\Phi}{\mathrm{d}u}\right)^2
-2\xi\Phi\frac{\mathrm{d}^2\Phi}{\mathrm{d}u^2}\right]y^2\nonumber\\
&&{}+F_1(u)y+F_0(u),
\end{eqnarray}
from which the functions $F_0$ and $F_1$ can be eliminated, and we
end with a configuration given by
\begin{subequations}\label{eq:TMGfreesol}
\begin{eqnarray}
ds^2&=&-\left\{F_2(u)\exp\!\left[-\mu(1-\kappa\xi\Phi^2)y\right]
+\frac{\kappa}{1-\kappa\xi\Phi^2}
\left[(1-2\xi)\left(\frac{\mathrm{d}\Phi}{\mathrm{d}u}\right)^2
-2\xi\Phi\frac{\mathrm{d}^2\Phi}{\mathrm{d}u^2}\right]y^2\right\}du^2
\nonumber\\
&&{}-2dudv+dy^2,\\
\Phi&=&\Phi(u).
\end{eqnarray}
\end{subequations}

As a first constatation, the inclusion of the topological mass has
broken the plane wave behavior of the gravitational field present in
$2+1$ gravity for a free field (\ref{eq:freesol}). In fact, the term
$\mu_{\mathrm{eff}}=\mu(1-\kappa\xi\Phi^2)$ acts as an effective
topological mass since the corresponding metric function $F$
satisfies
\begin{equation}\label{eq:K_Gmueff}
\Box{F}={\mu_{\mathrm{eff}}}^2F+\ldots,
\end{equation}
i.e., the Klein-Gordon equation with a retarded-time-dependent mass
term plus corrective terms due to the presence of the scalar source.
The above effective topological mass reduces to the vacuum value
$\mu_{\mathrm{eff}}=\mu$ for minimal coupling $\xi=0$. Indeed, in
the minimal case the gravitational field becomes
\begin{equation}\label{eq:TMGminsol}
ds^2=-\left[F_2(u)\mathrm{e}^{-\mu{y}}
+\kappa\left(\frac{\mathrm{d}\Phi}{\mathrm{d}u}\right)^2
y^2\right]du^2-2dudv+dy^2,
\end{equation}
and consists in a simple superposition of the vacuum \emph{pp} waves
of topologically massive gravity (\ref{eq:TMGppv}) and the $2+1$
gravity planes waves (\ref{eq:minsol}) with a free minimally coupled
scalar field as source. This is related to the fact that the
equation (\ref{eq:TMGuu}) is a inhomogeneous linear differential
equation whose solution is the sum of a particular solution,
determined from the scalar inhomogeneity, with a solution of the
corresponding homogenous system. Obviously, in absence of a scalar
field ($\Phi=0$) we recover from the above solution or from the
expression (\ref{eq:TMGfreesol}), in presence of nonminimal
coupling, the vacuum massive \emph{pp} waves (\ref{eq:TMGppv}).

For an huge value of the topological mass ($\mu\gg1$), the Einstein
tensor is dominant and one would expect that at this limit the
solutions (\ref{eq:TMGfreesol}) become the plane waves
(\ref{eq:freesol}) of $2+1$ gravity. This is indeed the case
provided the quantity $(1-\kappa\xi\Phi^2)$ to be positive. This
assumption is not too restrictive since in presence of nonminimal
coupling to gravity the term
$\kappa_{\mathrm{eff}}=\kappa/(1-\kappa\xi\Phi^2)$ acts as an
effective gravitational constant, and it must be positive at least
in the weak gravity limit in order to recover the intuitive
attractive behavior of gravity. On the light of this, under this
assumption the solution (\ref{eq:TMGfreesol}) can also be
interpreted as a superposition of the vacuum \emph{pp} waves of
topologically massive gravity (\ref{eq:TMGppv}) taking as
topological mass the effective one,
$\mu_{\mathrm{eff}}=\mu(1-\kappa\xi\Phi^2)$, together with the $2+1$
gravity plane waves (\ref{eq:freesol}) which have a free
nonminimally coupled scalar field as source.

In the next section, we shall see that the solutions described here
at a special limit when the topological mass goes to zero describe
new \emph{pp} wave solutions of conformal gravity.
\section{\label{sec:CG}Conformal Gravity \emph{pp} waves}

\subsection{\label{subsec:TMG2CG}Free conformal gravity \emph{pp} waves
from topologically massive gravity}

At the limit of small topological mass $\mu\to 0$ with
$\mu\kappa\sim 1$, the topologically massive gravity equations
(\ref{eq:TMG}) become \cite{Deser:1992nk}
\begin{eqnarray}\label{eq:CG}
C_{\alpha\beta}=\tilde{\kappa}T_{\alpha\beta},
\end{eqnarray}
where $\tilde{\kappa}$ is a dimensionless constant. Thus, at this
special limit, the contribution of the Einstein tensor disappears
and gravity is only governed by the Cotton tensor. At the first
sight, since the Cotton tensor is conformally invariant, this
equation (\ref{eq:CG}) seems to restrict the source to be also
conformally invariant. Two recent works have been done in this
direction where \emph{pp} waves configurations have been obtained
for a free conformal scalar field \cite{Deser:2004wd} and later
extended by the inclusion of a potential that does not spoil the
conformal invariance \cite{Ayon-Beato:2004fq}. In both cases, this
corresponds to choose the conformal coupling parameter $\xi=1/8$.
However, it appears that the solutions derived here for
topologically massive gravity (\ref{eq:TMGfreesol}) at the limit
discussed above generate new solutions of conformal gravity
equations (\ref{eq:CG}) for any arbitrary nonvanishing value of the
nonminimal coupling parameter $\xi$. Indeed, at the limit $\mu\to 0$
with $\mu\kappa\sim 1$, the solutions (\ref{eq:TMGfreesol}) become
\begin{subequations}\label{eq:CGfreesol}
\begin{eqnarray}
ds^2&=&-\left[F_2(u)\mathrm{e}^{\tilde{\kappa}\xi\Phi^2y}
+2\Phi^{\frac{1-4\xi}{2\xi}}\frac{\mathrm{d}}{\mathrm{d}u}\left(
\Phi^{\frac{2\xi-1}{2\xi}}\frac{\mathrm{d}\Phi}{\mathrm{d}u}\right)y^2
\right]du^2
-2dudv+dy^2,\\
\Phi&=&\Phi(u),
\end{eqnarray}
\end{subequations}
and, a simple check shows that these limiting configurations are
effectively solutions of conformal gravity (\ref{eq:CG}). In
particular, for the conformal coupling $\xi=1/8$, we recover exactly
the \emph{pp} wave solutions of Ref.~\cite{Deser:2004wd}. For
$\xi\not=1/8$, the matter source is not \emph{a priori} conformally
invariant and so a natural question is to ask why these limiting
configurations are solutions of conformal gravity. The
``miraculous'' lies in the fact that \emph{pp} wave backgrounds
impose to the matter to have only one nonvanishing component of the
energy-momentum tensor, namely $T_{uu}$, and since $g^{uu}=0$, the
energy-momentum tensor is traceless on-shell. In other words, the
conformal character of the equations (\ref{eq:CG}) is preserved
on-shell. Various questions emerge from the last observation. For
example, one can asks whether the limiting solutions
(\ref{eq:CGfreesol}) are the only ones of conformal gravity with a
free source for any value of the nonminimal coupling parameter or if
their exist other configurations. A simple calculation shows that
for $\xi\ne0$ the limiting configurations (\ref{eq:CGfreesol}) are
the only \emph{pp} wave solutions of conformal gravity equations
(\ref{eq:CG}). Since the minimal coupling limit is singular in
Eq.~(\ref{eq:CGfreesol}), we solve independently the original
equations (\ref{eq:CG}) in the case of the minimal coupling $\xi=0$
and we get the following solution
\begin{subequations}\label{eq:CGfreesolxi=0}
\begin{eqnarray}
ds^2&=&-\left(\frac{\tilde{\kappa}}{3}
\left(\frac{\mathrm{d}\Phi}{\mathrm{d}u}\right)^2y^3
+F_2(u)y^2\right)du^2-2dudv+dy^2,\\
\Phi&=&\Phi(u).
\end{eqnarray}
\end{subequations}
Note that in absence of source (i.e., $\Phi=0$) one recovers the
plane wave solution of the vacuum conformal gravity.

The argument which consists of looking for configurations that have
a traceless energy-momentum tensor on-shell can be applied for any
matter source (not necessarily a free scalar field). In particular,
in what follows, we consider the self-interacting case for an
arbitrary value of the nonminimal coupling parameter.

\subsection{\label{subsec:CGU<>0}Self-interacting conformal gravity
\emph{pp} waves}

Motivated by the previous study, we explore the existence of
\emph{pp} wave configurations for conformal gravity with a
self-interacting source and for a generic value of the nonminimal
coupling parameter. Due to the fact that for a \emph{pp} wave
ansatz, $C_{uu}$ is the only nonvanishing component of the Cotton
tensor, the arguments applied in Sec.~\ref{sec:U<>0} are still
valid. In particular, the potential must have the form given by the
expression (\ref{eq:U(Phi)}) while the scalar field is expressed by
Eqs.~(\ref{eq:Phi2sigma}) and (\ref{eq:subssigma}). Making the
following redefinitions of the involved variables
\begin{subequations}\label{eq:xH2yFcg}
\begin{eqnarray}
x&=&\frac{\tilde{\kappa}\xi(1-4\xi)}{2\sqrt{\lambda}(1-8\xi)}
\left(2\sqrt{\lambda}y+f\right)^{(1-8\xi)/(1-4\xi)}, \\
H(u,x)&=& \frac{F}{2\sqrt{\lambda}y+f}-\frac{1}{2\lambda}
\frac{\mathrm{d}^2f}{\mathrm{d}u^2},
\end{eqnarray}
\end{subequations}
the component $uu$ of the equations (\ref{eq:CG}) can be written as
\begin{equation}\label{eq:ghypergeomDE}
x^2\partial^3_{xxx}H-x(x-3)\partial^2_{xx}H
-2\frac{(1-8\xi)^2x+4\xi(1-6\xi)}{(1-8\xi)^2}\partial_xH
-\frac{4\xi(1-4\xi)}{(1-8\xi)^2}H=0.
\end{equation}
The above equation is the generalized hypergeometric differential
equation \cite{Erdelyi:1953} for which the general solution reads
\begin{eqnarray}\label{eq:solHcg}
H(u,x)&=&F_1(u)\,_1\!\tilde{F}_1\!\left(\frac{1-4\xi}{1-8\xi};
                            \frac{2(1-6\xi)}{1-8\xi};x\right)
        +F_2(u)\left(\frac{2\sqrt{\lambda}(1-8\xi)}
         {\tilde{\kappa}\xi(1-4\xi)}x\right)^{-\frac{1-4\xi}{1-8\xi}}
                            \nonumber\\
      & &{}+F_3(u)\left(\frac{2\sqrt{\lambda}(1-8\xi)}
        {\tilde{\kappa}\xi(1-4\xi)}x\right)^{\frac{1-4\xi}{1-8\xi}}
        \,_2\!\tilde{F}_2\!\left(1,\frac{2(1-4\xi)}{1-8\xi};
        \frac{3-16\xi}{1-8\xi},\frac{2(1-6\xi)}{1-8\xi};x\right).
        \qquad~
\end{eqnarray}
Here $_1\!\tilde{F}_1(a;b;x)$ and $_2\!\tilde{F}_2(a,b;c,d;x)$
denote the corresponding generalized hypergeometric functions
\cite{Erdelyi:1953}. Returning to the original variables by means of
Eqs.~(\ref{eq:xH2yFcg}) and after making the coordinate change
(\ref{eq:coordppw}) we obtain the final solution
\begin{subequations}\label{eq:solcg}
\begin{eqnarray}
ds^2&=&-\biggl[F_1(u)\,_1\!\tilde{F}_1\!\left(\frac{1-4\xi}{1-8\xi};
                            \frac{2(1-6\xi)}{1-8\xi};
\frac{\tilde{\kappa}\xi(1-4\xi)}{2\sqrt{\lambda}(1-8\xi)}
\Phi^{\frac{8\xi-1}{2\xi}}\right)
       2\sqrt{\lambda}y\nonumber\\
    & &{}+F_3(u)\,_2\!\tilde{F}_2\!\left(1,\frac{2(1-4\xi)}{1-8\xi};
        \frac{3-16\xi}{1-8\xi},\frac{2(1-6\xi)}{1-8\xi};
    \frac{\tilde{\kappa}\xi(1-4\xi)}{2\sqrt{\lambda}(1-8\xi)}
\Phi^{\frac{8\xi-1}{2\xi}}\right)
        4\lambda{y}^2\biggr]du^2
        \nonumber\\
    & &{}-2dudv+dy^2,\qquad~\\
\Phi&=&(2\sqrt{\lambda}y)^{-2\xi/(1-4\xi)}.
\end{eqnarray}
\end{subequations}
We shall not intent here to cover the singular values of the
nonminimal coupling parameter in the above solution. This can be
done along the same line than in the self-interacting $2+1$ gravity
case of Appendix \ref{sec:xi_n}. We just notice that the conformal
coupling $\xi=1/8$ belongs to those singular values but this case
has already been derived in Ref.~\cite{Ayon-Beato:2004fq}.

\section{\label{sec:conclu}Conclusions}

As it is well-known, pure gravitational waves are forbidden in three
dimensions. A natural way to circumvent this problem is to consider
a matter source as it has been done here for a scalar field
nonminimally coupled to a \emph{pp} wave with or without
self-interaction potential. For Einstein gravity, in the free case,
we obtain gravitational plane wave solutions for which the scalar
field also behaves like a wave. Additionally, the gravitational wave
profile is fixed in terms of the scalar one.

The introduction of a self-interaction potential has several
consequences. Firstly, its form is dictated by the field equations
which only allow power-law potentials with powers given in terms of
the nonminimal coupling parameter. For the conformal coupling, this
potential reduces exactly to the conformally invariant one in three
dimensions. Secondly, the presence of the self-interaction breaks
explicitly the wavy behavior of the scalar field that was present in
the free case. Indeed, the scalar field loses its arbitrary
dependence on the retarded time, and is now an explicit function of
the wave-front coordinate. For the special value of the coupling
parameter $\xi=1/2$, the corresponding potential becomes a positive
constant, and hence the related configuration can be seen as a free
scalar field that solve the Einstein equations in presence of an
effective positive cosmological constant. The special value
$\xi=1/4$, studied separately in Appendix \ref{app:xi=1/4}, is also
interpreted in a different way. In this case the potential reduces
to a mass term, and consequently the system describes a free massive
scalar field with the corresponding nonminimal coupling to gravity.

We have also considered the natural extension for which \emph{pp}
waves are rigged by topologically massive gravity with a free
nonminimally coupled scalar field acting as a source. In the
particular case of minimal coupling, the solutions can be seen as a
simple superposition of the vacuum \emph{pp} waves of topologically
massive gravity and the Einstein gravity plane waves with a free
scalar source. In the nonminimal case, the solutions can be
interpreted again as a superposition of the corresponding solutions
of topologically massive gravity and Einstein gravity, provided the
topological mass is changed by an effective one depending on the
nonminimal coupling.

At the small topological mass limit with a huge gravitational
constant, new solutions of conformal gravity with a free source have
been obtained for any arbitrary value of the nonminimal coupling
parameter. This is not in contradiction with the conformal character
of conformal gravity since these limiting solutions have a traceless
energy-momentum tensor on-shell. This fact is due to the particular
\emph{pp} wave ansatz that restricts the matter source to have only
one nonvanishing energy-momentum tensor component along the retarded
time. Motivated by the results of the above limit, we have also
extended the previous nonminimally coupled configurations to the
self-interacting case. Interestingly, the allowed potentials are
exactly the same than those arising in the case of Einstein gravity.

It would be interesting to explore the existence of other background
geometries that allow special superposition of solutions as those
arising here. From this study, it is also natural to consider other
matter source to conformal gravity that preserves the conformal
invariance on-shell. Another interesting work will consist to extend
these considerations in arbitrary $D$ dimension and to see whether
these features are still valid or were only specific to the
three-dimensional case. In this case, an interesting option would be
to consider front wave geometries which are not necessarily planes.

\begin{acknowledgments}
We thank Roman Jackiw and Jorge Zanelli for useful discussions. This
work is partially supported by grants 3020032, 1040921, 7040190, 1051064 and 1051084 from FONDECYT, grants 38495E and 34222E from CONACyT, grant
2001-5-02-159 from CONICYT/CONACyT, and grant D-13775 from
Fundaci\'on Andes. Institutional support to the Centro de Estudios
Cient\'{\i}ficos (CECS) from Empresas CMPC is gratefully
acknowledged. CECS is a Millennium Science Institute and is funded
in part by grants from Fundaci\'{o}n Andes and the Tinker
Foundation.
\end{acknowledgments}

\appendix

\section{\label{app:xi=1/4}\emph{pp} waves from nonminimal coupling
$\xi=1/4$}

For the specific value $\xi=1/4$ of the nonminimal coupling
parameter, our substitution (\ref{eq:Phi2sigma}) for a
self-interacting scalar field is not valid and instead we consider
\begin{equation}\label{eq:Phi2sigma1/4}
\Phi=\frac1{\sqrt{\kappa}}\mathrm{e}^{\sigma}.
\end{equation}
With this substitution the function $\sigma$ satisfy the same
equations than in the generic case, i.e., Eqs.~(\ref{eq:s_yyuy}),
and is given by
\begin{equation}\label{eq:subssigma1/4}
\sigma(u,y)=my+f(u),
\end{equation}
while the allowed potential is now a simple mass term
\begin{equation}\label{eq:U1/4(Phi)}
U_{1/4}(\Phi)=\frac12m^2\Phi^2.
\end{equation}
The remaining Einstein equation (\ref{eq:uu-uv}) is expressed as
\begin{equation}
\left(4\mathrm{e}^{-2(my+f)}-1\right)\partial^2_{yy}F-2m\partial_yF
+4\frac{\mathrm{d}^2f}{\mathrm{d}u^2}=0,
\end{equation}
whose solution reads
\begin{equation}
F(u,y)=\left(F_1(u)+\frac1{m^2}\frac{\mathrm{d}^2f}{\mathrm{d}u^2}\right)
\ln{\left(4\mathrm{e}^{-2(my+f)}-1\right)}
+\frac2{m^2}\frac{\mathrm{d}^2f}{\mathrm{d}u^2}(my+f)+F_2(u).
\end{equation}
After performing the coordinate transformation (\ref{eq:coordpw}),
we finally obtain the following solution
\begin{eqnarray}\label{eq:sol1/4}
ds^2&=&-F_1(u)\ln{\left(4\mathrm{e}^{-2my}-1\right)}du^2-2dudv+dy^2,\qquad~\\
\Phi&=&\frac{1}{\sqrt{\kappa}}\mathrm{e}^{my},
\end{eqnarray}
which describe a free massive scalar field nonminimally coupled to a
\emph{pp} wave with parameter $\xi=1/4$.

\section{\label{sec:xi_n}\emph{pp} waves from nonminimal couplings
$\xi_n=1/[4(2+n)]$}

Here we analyze the singular cases not covered within the
self-interacting solution (\ref{eq:solH}) for Einstein gravity. The
hypergeometric function $_2\!\tilde{F}_1(a,b;c;x)$ is not defined
when the parameter $c$ is equal to a non-positive integer $-m$
provided that $a$ or $b$ is not equal to a negative integer $-n$
with $n<m$. In our case, solution (\ref{eq:solH}) is of the form
$_2\!\tilde{F}_1(1,b;b+1;x)$ where $b=(4\xi-1)/(4\xi)$ and
consequently, we must exclude from the previous analysis the
following values of the nonminimal coupling parameter
\begin{equation}\label{eq:xi_n}
\xi_n=\frac{1}{4(n+2)}, \qquad \mathrm{where} \quad n\in\mathbb{N},
\end{equation}
which includes in particular the conformal coupling $\xi_0=1/8$. The
self-interaction potentials corresponding to these special
nonminimal couplings are
\begin{equation}\label{eq:Un(Phi)}
U_n(\Phi)=\frac{\lambda}{2(n+1)^2}\Phi^{2(2n+3)}.
\end{equation}

We now analyze the \emph{pp} waves configurations for these
particular values. For $\xi=\xi_n$, the integration of equation
(\ref{eq:hypergeomDE}) for each $n$ leads to the following solution
\begin{equation}\label{eq:H_n}
H_n(x)=\left\{F_1(u)\left[\ln\left(1-\frac{1}{x}\right)
+\sum_{l=1}^{n+1}\frac{1}{lx^l}\right]+F_2(u)\right\}
\left(\frac{x}{\kappa\xi_n}\right)^{n+1}.
\end{equation}
Recovering the original variables from definitions (\ref{eq:xH2yF})
and using again the coordinate transformation (\ref{eq:coordppw})
the final solution for the discrete values of the nonminimal
coupling parameter $\xi_n=1/[4(2+n)]$, $n=0,1,2,\ldots$, reads
\begin{subequations}\label{eq:soln}
\begin{eqnarray}
ds^2&=&-F_1(u)\left[\ln\left(1-\frac{1}{\kappa\xi_n\Phi^2}\right)
+\sum_{l=1}^{n+1}\frac{1}{l(\kappa\xi_n\Phi^2)^l}\right]
du^2-2dudv+dy^2,\qquad~\\
\Phi&=&\frac1{(2\sqrt{\lambda}y)^{1/[2(n+1)]}}.
\end{eqnarray}
\end{subequations}
In particular, when the source is conformally invariant [i.e.,
$\xi_0=1/8$ with the conformal potential (\ref{eq:U1/8(Phi)})] the
above expression reduces to
\begin{subequations}\label{eq:sol1/8}
\begin{eqnarray}
ds^2&=&-F_1(u)\left[\ln\left(1-\frac{16\sqrt{\lambda}y}{\kappa}\right)
+\frac{16\sqrt{\lambda}y}{\kappa}\right]
du^2-2dudv+dy^2,\\
\Phi&=&\frac1{\sqrt{2\sqrt{\lambda}y}}.
\end{eqnarray}
\end{subequations}


\end{document}